\documentclass[a4paper]{article}

\usepackage{INTERSPEECH2022}

\usepackage{color,soul}

\usepackage{hyperref}
\usepackage{cleveref}

\usepackage{arydshln}

\usepackage{caption}
\usepackage{subcaption}

\usepackage{booktabs}
\usepackage[table,xcdraw]{xcolor}

\title{Speech and the n-Back task as a lens into depression. How combining both may allow us to isolate different core symptoms of depression.}

\name{Salvatore Fara$^1$, Stefano Goria $^1$, Emilia Molimpakis$^1$, Nicholas Cummins$^{1,2}$ \thanks{Submitted to Interspeech 2022}}

\address{
  $^1$Thymia, London , UK\\
  $^2$Institute of Psychiatry, Psychology \& Neuroscience (IoPPN), King’s College London, London, UK}
\email{\{salvatore,stefano,emilia,nick\}@thymia.ai, nick.cummins@kcl.ac.uk}

\begin{document}

\maketitle

\begin{abstract}
Embedded in any speech signal is a rich combination of cognitive, neuromuscular and physiological information. This richness makes speech a powerful signal in relation to a range of different health conditions, including major depressive disorders (MDD). One pivotal issue in speech-depression research is the assumption that depressive severity is the dominant measurable effect. However, given the heterogeneous clinical profile of MDD, it may actually be the case that speech alterations are more strongly associated with subsets of key depression symptoms. This paper presents strong evidence in support of this argument. First, we present a novel large, cross-sectional, multi-modal dataset collected at Thymia. We then present a set of machine learning experiments that demonstrate that combining speech with features from an n-Back working memory assessment improves classifier performance when predicting the popular eight-item Patient Health Questionnaire depression scale (PHQ-8). Finally, we present a set of experiments that highlight the association between different speech and n-Back markers at the PHQ-8 item level. Specifically, we observe that somatic and psychomotor symptoms are more strongly associated with n-Back performance scores, whilst the other items: anhedonia, depressed mood, change in appetite, feelings of worthlessness and trouble concentrating are more strongly associated with speech changes.

\end{abstract}


\section{Introduction}


Major depressive disorders (MDDs) are a constantly growing economic and societal problem~\cite{OECD2018}. The harsh collateral social side-effects of COVID-19 (social isolation, employment loss, bereavement, grief) have further exacerbated this already substantial problem~\cite{moreno2020mental, ons2021corona}. MDDs are also one of the leading causes of disability worldwide, accompanied by high socio-economic costs~\cite{deloitte2020cost}. As a result, accurately identifying, treating and thereby reducing the prevalence of MDDs is a major public health goal and challenge. However, globally, demand for mental health support greatly outstrips supply.Therefore, advances in digital health tools and phenotyping technologies that can support clinicians in this effort are crucial to ensure better and more widespread access to high-quality mental health support services and treatment.


Speech has great potential to be a source of such phenotypes and to provide unique preventative and predictive information about depression~\cite{cummins2015review,low2020review, yamamoto2020using, abbas2021remote}. However, current research in this space is not without its limitations. These become apparent when we consider a) the datasets used, b) core assumptions regarding effects of depression severity, and c) additional modalities collected alongside speech.

If we look at machine learning approaches that have arguably dominated speech-depression research over the last ten years, the majority of these works use the Audio/Visual Emotion Challenge (AVEC) datasets~\cite{valstar2013avec,Ringeval2017avec}. Whilst these investigations have undoubtedly advanced our knowledge in modelling depressed speech, their continuous use in pseudo-competition settings raises concerns relating to Goodhart’s Law~\cite{Miller2020} and overfitting~\cite{hawkins2004problem}. Therefore, in order for the speech-depression community to continue to be driven forward, it is imperative that new databases are (additionally) used henceforth.

Another critical issue in speech-depression research is that the majority of works assume depression severity is in, and of itself, the most dominant and important measurable effect in speech. However, depression has a highly heterogeneous clinical profile; it may be the case therefore, that speech alterations observed within depression could be more strongly (and more explainably) associated with subsets of core depressive symptoms, rather than severity as an absolute measure. Preliminary works exploring this conjecture have indeed demonstrated that speaking rate measures are more strongly correlated with mood and psychomotor retardation measurements than with overall depression severity~\cite{Horwitz2013Importance, trevino2011phonologically}. These findings, however, are from small sample sizes and should be regarded as preliminary only. 



Finally, no research to date has considered placing depression assessment via speech - whether with respect to depressive severity or depressive symptoms - alongside assessments of depression from other modalities; other than vision~\cite{cohn2018multimodal, girard2015automated, pampouchidou2017automatic}. For example, to the best of the authors' knowledge, there is no research examining how speech alterations may be complemented with reaction times and error rates in more classic neuropsychology protocols~\cite{nikolin2021investigation, owen2005n}. If we are to attempt to understand how the speech signal may be related to depressive symptomatology, it makes sense to examine it alongside and within the context of such classic protocols, leveraging complementary information to gain a better understanding of this relationship.


In response to the limitations described above, this paper presents a set of preliminary analyses conducted on a novel, large, online multimodal dataset collected by Thymia Limited (henceforth Thymia). In the following sections, we detail how for the first time, we combined speech elicitation tasks with the n-Back Task, an experimental psychology protocol targeting Working Memory~\cite{nikolin2021investigation, 10.3389/fpsyg.2019.00004, ROSE2006149}. Presented analyses include a set of machine learning experiments that highlight the complementary information contained within the speech and n-Back features when predicting the presence or absence of depression. Importantly, we additionally present a set of experiments that highlight the association between different speech and n-Back markers at the symptom level of depression, allowing us an initial insight into how speech and the n-Back Task may be used in tandem to better target different core depressive symptoms.

\section{Data Collection}
As part of its core mission Thymia \cite{thymiawebsite} (a London based mental health tech startup) is actively collecting 
multimodal -- video, speech and behavioural -- data to develop models targeting remote assessment and monitoring of depression. In order to maximise access to the studies, the data collection needs to run on a browser, operating system and device agnostic platform. 

\subsection{Thymia Research Platform}

In the past few years several platforms for online studies have gained popularity \cite{research-internet}, but to the best of our knowledge none offer the level of flexibility and security required for our intended task. Thymia, therefore, developed and implemented our own research platform.

The Thymia Research platform allows the hosting of complex, remote, one-off or longitudinal multimodal studies where detailed, informed consent and demographic data can be gathered, questionnaires can be completed, and gamified activities can be assigned to participants on a schedule. During these activities, data from the device’s camera, keyboard, mouse/trackpad and/or touch screen can be streamed to a secure backend\footnote{The Thymia platform is fully compliant with the 2018 EU General Data Protection Regulations, is ISO27001-certified and NHS Toolkit-compliant. All of our research is reviewed by independent and/or University research ethics committees. The Thymia platform has been successfully used not only by Thymia, but also by several UK university research groups.}. Throughout the experiment protocols, when launching activities that require media recording, participants are reminded that their camera and/or microphone will be switched on for recording and they are free to opt out. 

\subsection{Online Study Setup}
The dataset used in this work is part of a larger online study\footnote{“Does mood affect speech patterns and reactions? A Proof of Concept study.” This study, including all Information Sheets and Consent Forms, has been reviewed by an independent research ethics expert working under the auspices of the Association of Research Managers and Administrators. All subjects read a detailed Information Sheet prior to beginning the experiment, could remotely ask questions and expressly consented to participate knowing they were free to withdraw at any point. All data collected were handled according to GDPR and participants were compensated for their time.} running on the Thymia Research platform. It consists of demographic and psychiatric questionnaires, speech eliciting tasks and gamified experimental protocols targeting visual processing, attention, psychomotor response and working memory. 
Participants were pre-screened and split into two groups: a patient and a healthy age- and gender-matched control group. Both groups consisted of adult, native English speakers, aged 18 to 75 (evenly split across age groups), with normal or corrected-to-normal vision, no hearing, language or speech impairments and - for the control group - no prior history of psychiatric illness. The patient group participants must have had a formal MDD diagnosis by a GP, clinical psychologist or psychiatrist at least two months prior to participating.

A study session is completed via the participant’s laptop or smart device without any researcher supervision. The session includes standardised questionnaires to gather information about demographics and mood, including the \textit{Patient Health Questionnaire - 8} (PHQ-8)~\cite{kroenke2009phq}, a well-established depression scale used commonly in research which aims to assess a number of core depressive symptoms, including fatigue, working memory impairment, anhedonia and low mood. Participants also completed speech eliciting tasks, including an Image Description Task, and short point-and-click (or screen tapping) tasks measuring reaction times, accuracy and error rates including an implementation of the classical n-Back task~\cite{owen2005n}.


\begin{figure}[t]
  \centering
  \includegraphics[width=0.95\linewidth]{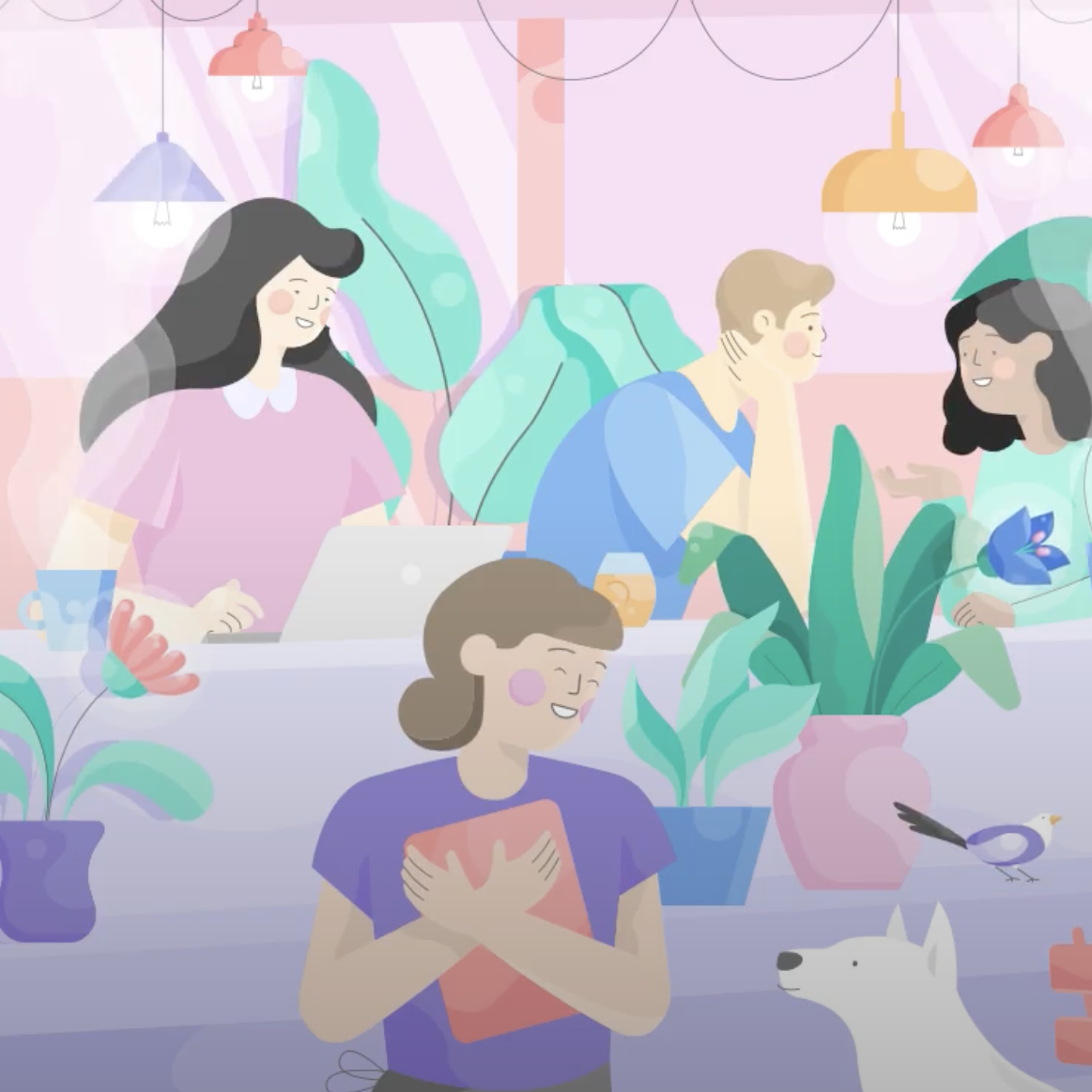}
  \caption{
  The Image Description Task; the image represents a cafe' scene with several people. The image has many elements that are animated: the bird is moving, people are talking.}
  \label{fig:image_description}
  \vspace{-5mm}
\end{figure}

\subsubsection{Image Description Task}
In the Image Description Task, participants are encouraged to describe what they see in the image whilst keeping their camera and microphone on; while performing the task participants can see their own camera feed as feedback and reminder of the fact that they are being recorded. The image itself is a rich, animated illustration depicting a caf\'{e} environment filled with people at different tables (Figure~\ref{fig:image_description}).



\subsubsection{n-Back Task}
Performance in the n-Back Task offers insights into working memory dysfunction in depression~\cite{nikolin2021investigation, ROSE2006149}. The implementation used as part of this study is presented as a card memory game. A rounded edged card (as in a deck of cards) with a single digit number or letter in the middle of it appears in the centre of a blank screen for a short interval; this is then followed by a blank screen; followed by another card containing another number or letter; followed by a blank screen etc. The participant must tap/click the screen when the current number or letter matches the number or letter that appeared \textit{n} cards back. 

There are two difficulty levels based on the value of \textit{n}. The different difficulty levels are based on progressive cognitive loading. In the first block, \textit{n} equals one (1) (i.e. the current target card must match the card one card back); in the second block \textit{n} increases to two (2) (i.e. the current target card matches the card appearing two cards before it). Each participant saw 3 practice blocks for each n-Back load, followed by 6 experiment blocks for each n-Back load. Match number and position were pseudo-randomised across blocks and block order was counterbalanced across participants. 

\begin{table*}[t!]
  \caption{Sociodemographic, Depression (Low: PHQ-8 $<$ 10, High: PHQ-8 $\geq$ 10) and Activity distributions in the experimental data. 
  }
  \vspace{-2mm}
  \label{tab:demographics}
  \centering
  \begin{tabular}{c c c c c c c c c}
    \toprule
    \multicolumn{1}{c}{\textbf{Partition}} &
    \multicolumn{2}{c}{\textbf{\#Participants}} &
    \multicolumn{2}{c}{\textbf{Age}} &
    \multicolumn{2}{c}{\textbf{Binary PHQ-8}} & 
    \multicolumn{2}{c}{\textbf{Total Activity Time}} \\
    \multicolumn{1}{c}{\textbf{}} &
    \multicolumn{1}{c}{\textbf{Male}} &
    \multicolumn{1}{c}{\textbf{Female}} &
    \multicolumn{1}{c}{\textbf{Mean}} &
    \multicolumn{1}{c}{\textbf{SD}} &
    \multicolumn{1}{c}{\textbf{\#Low}} &
    \multicolumn{1}{c}{\textbf{\#High}} &
    \multicolumn{1}{c}{\textbf{Speech}} & 
    \multicolumn{1}{c}{\textbf{n-Back}} \\
    \midrule
    Training & 387 & 388 & 34.94 & 12.64 & 501 & 274 & 10:19:36 & 4 days, 17:13:01 \\ 
    Test & 97 & 97 & 35.39 & 13.24 & 125 & 69 & 2:21:33 & 1 day, 2:12:18 \\ 
    \bottomrule
  \end{tabular}
   \vspace{-2mm}
\end{table*}

\section{Dataset}



Our experimental dataset consists of 969 participants who performed a range of activities within a single session on the Thymia Research Platform using their own personal devices. The presented analysis focuses on two specific data modalities gathered through the platform: audio data from the Image Description Task, and behavioural data from the n-Back Task. 

We performed a stratified split of the dataset into training and test sets (80/20 \%), keeping the same proportions of genders, age groups and PHQ-8 distribution (Table~\ref{tab:demographics}).

\subsection{Quality Controls}

Given the real-life nature of the dataset, each participant using their own personal device, we performed a number of quality controls on the different data modalities. 

\begin{figure}[t!]
  \centering
  \includegraphics[width=0.925\linewidth]{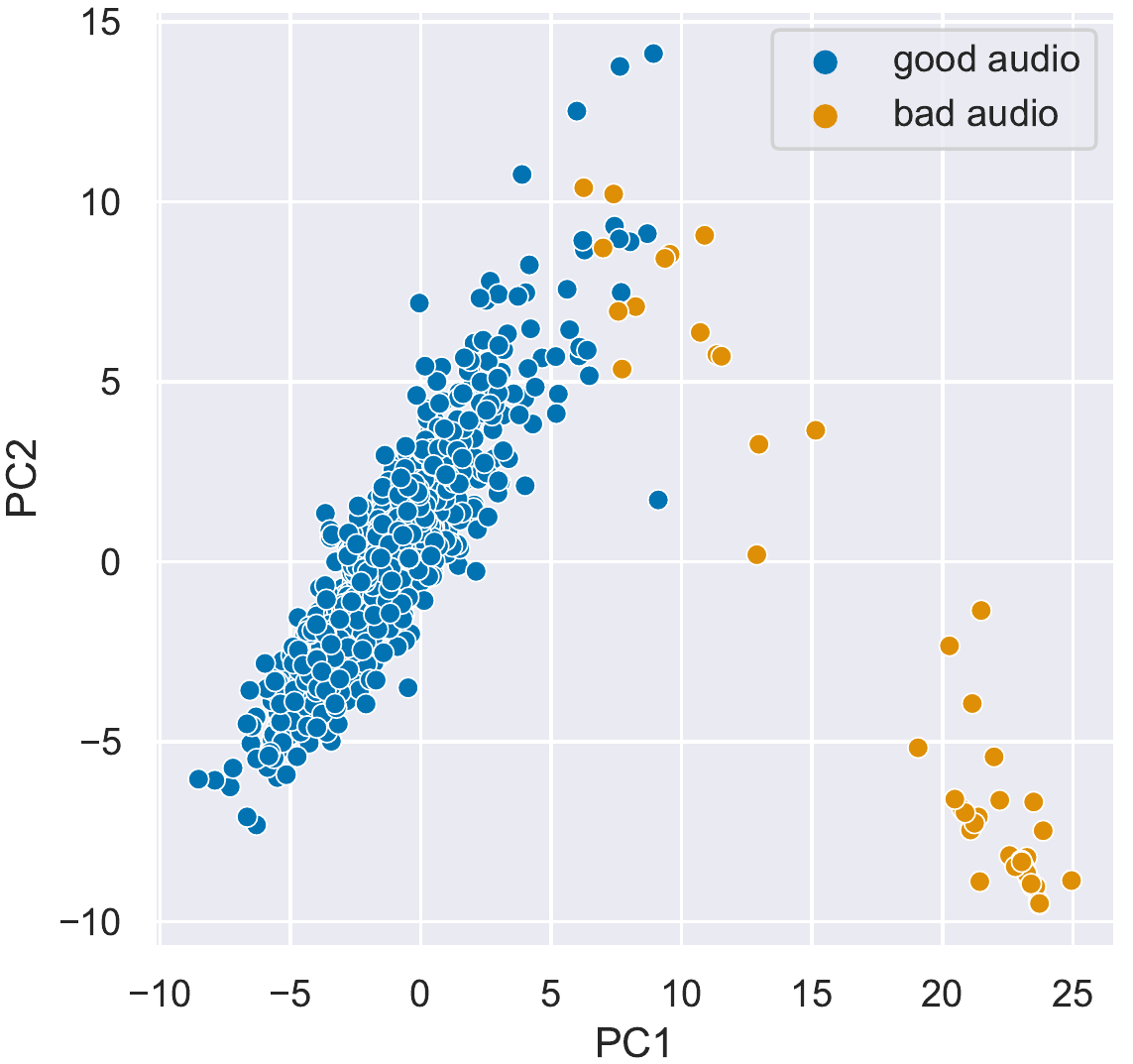}
  \caption{
  First and second principal components of the audio features from the Image Description Task highlighting the difference in distribution of good and bad audio files.}
  \label{fig:audio_quality}
   \vspace{-5mm}
\end{figure}

For the audio recordings, we implemented an automated audio quality pipeline to flag the audio files as having good or bad quality based on two criteria: (i) the quality of extracted audio features, and (ii) the presence of speech activity. Out of 917 files, the audio quality pipeline detected 51 bad quality audio files that were rejected from subsequent analysis. Through manual inspection of these 51 files, we confirmed that they were all cases of either high environmental noise, microphone turned-off or malfunctioning, or participant not speaking. A principal component projection of the extracted features of each audio file in the dataset, confirms the difference between the distributions of good and bad audio files (Figure~\ref{fig:audio_quality}). 

For the 
n-Back Task, we confirmed that the various metrics of performance are within expected ranges and vary with n-Back load as well as known covariates such as age; noting a general decrease in n-Back performance with age~\cite{10.3389/fpsyg.2018.02208}.




\section{Experimental Settings}

We investigated the impact of combining multiple data modalities, namely audio data from the Image Description Task and behavioural data from the n-Back Task, in a binary PHQ-8 classification paradigm (PHQ-8 $<$ 10 vs PHQ-8 $\geq$ 10). Details on the features and the models used are provided in the following.

\subsection{Speech Features}

The audio recordings of the Image Description Task were processed to extract a range of acoustic and linguistic features. We extracted 88 acoustic features as defined in the \textit{extended Geneva Minimalistic Acoustic Parameter Set} (eGeMAPS) \cite{eyben2015geneva} using 
 \textsc{openSMILE}~\cite{eyben2010opensmile}. In addition, we extracted a curated set of 28 features describing speech rate, pitch, voice quality and formant properties, using the Parselmouth package \cite{parselmouth} as a Python interface to Praat~\cite{boersma2001praat}. To quantify the linguistic content, 
 we first transcribed the 
 files using Amazon Transcribe~\cite{aws-transcribe}. We then used the spaCy library~\cite{spacy2} to extract 
 25 linguistic features describing speech-rate, pause-rate and part-of-speech usage.


\subsection{n-Back Features}

For each session, the data collected from the n-Back Task consists of a sequence of clicks and a corresponding sequence of targets and non-targets. From these sequences, we calculated standard features that quantify the performance in the task, namely precision, recall, false-positive rate, as well as reaction time. These features were calculated separately for the two n-Back loads (1-Back and 2-Back), yielding 
8 n-Back features. 

\subsection{Models}

All models, training and calibration procedures are implemented using the scikit-learn package~\cite{scikit-learn}.

\begin{figure}[b!]
  \centering
  \includegraphics[width=\linewidth]{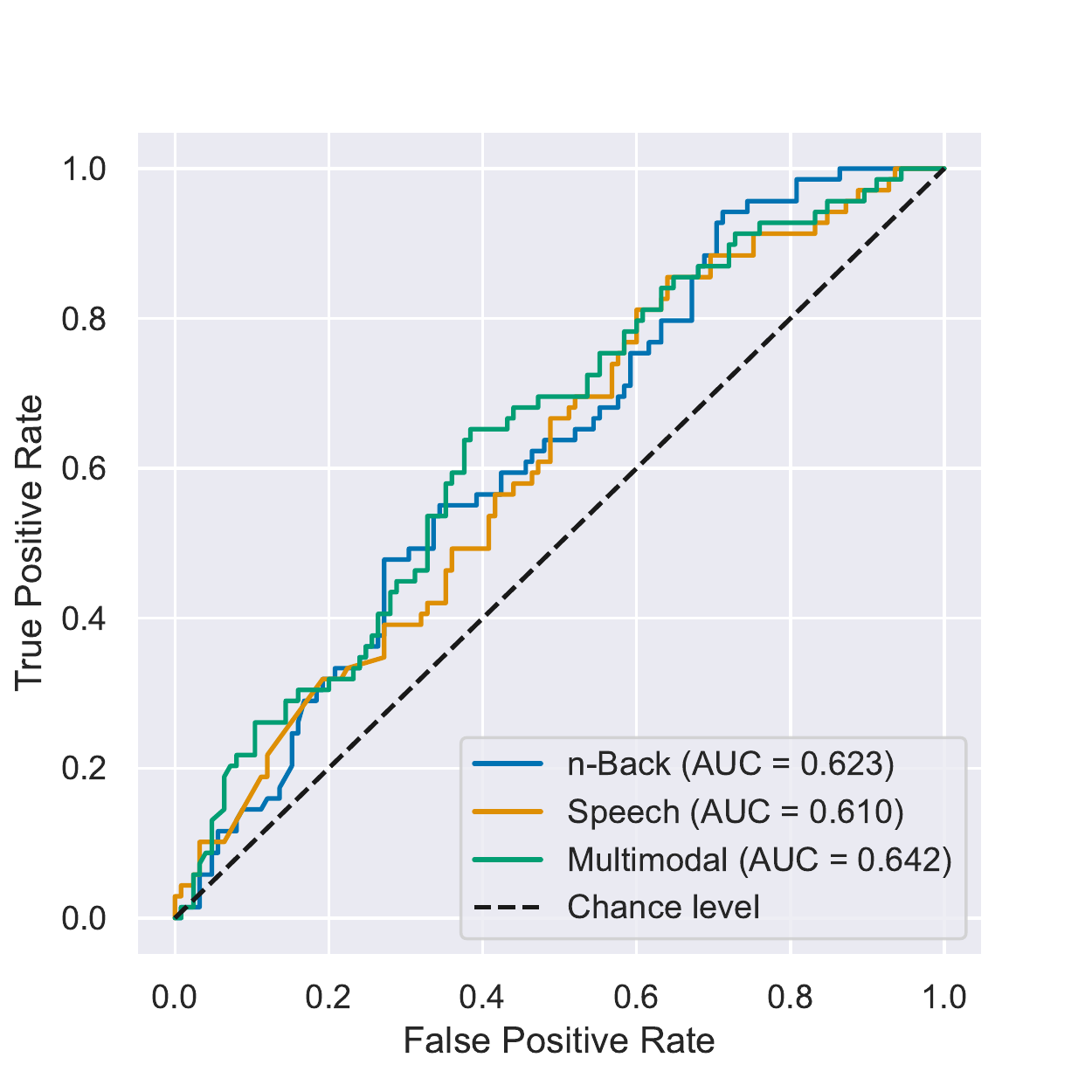}
  \caption{ROC curves calculated on the hold-out test set for the Speech, n-Back and Multimodal models.}
  \label{fig:test_roc}
\end{figure}

\subsubsection{Speech and n-Back Models}

First, we investigated five unimodal models: four speech models and an n-Back model. The speech models are; (i) an eGeMAPS  model; (ii) a Praat model; (iii) a Linguistic model; and (iv) a Speech model, which is the early fusion model of the 141 acoustic and linguistic features. All models consist of a binary Random Forest classifier with input features from their specific feature representations. All models receive 3 additional features as input, namely age, gender, and device setup. This last categorical feature encodes the personal device setup on which the session was performed (i.e. Laptop+Trackpad, Laptop+Mouse, Mobile/Tablet, Desktop). Standard rescaling is applied to all numerical features, while gender is encoded as binary and device setup is one-hot encoded. The unimodal models are fitted and calibrated on the training set using a cross-validated random parameter search with 100 iterations and 10 cross-validation folds. The hyperparameters tuned were \textit{\#Trees}, \textit{Max. Depth}, \textit{Min. \#Samples/Split} and \textit{Max. Rel. \#Features}.


\subsubsection{Multimodal Model}

We combined the Speech and n-Back unimodal models into a voting ensemble to create a multimodal model. The predictions of this model are given by a soft-voting rule, whereby the multimodal prediction is given by the most likely class label after averaging the predicted class probabilities across the unimodal models.

\begin{table}[]
  \caption{Cross-validated model performances when classifying low (PHQ-8 $<$ 10) versus high (PHQ-8 $\geq$ 10) depression on the training set}
  \label{tab:cv_performance}
  \centering
  \begin{tabular}{@{}l r r@{}}
    \toprule
    \multicolumn{1}{l}{\textbf{Model}} &
    \multicolumn{2}{c}{\textbf{ROC-AUC}}  \\
    \multicolumn{1}{c}{\textbf{}} &
    \multicolumn{1}{c}{\textbf{Mean}} & 
    \multicolumn{1}{c}{\textbf{SD}} \\
    \midrule
    eGeMAPS & 0.620 & 0.032 \\ 
    Praat & 0.607 & 0.056 \\ 
    Linguistic & 0.625 & 0.045 \\ 
    n-Back & 0.619 & 0.074 \\ 
    Speech & 0.631 & 0.024 \\ \midrule
    Multimodal & \textbf{0.652} & 0.037 \\ 
    \bottomrule
  \end{tabular}
\end{table}


\subsubsection{Feature Analysis}
To gain further insights into the performance of our multimodal system, we ran a set of linear regression analyses on our training set. The aim of this testing was to establish the importance of different eGeMAPs, Praat, Linguistic and n-Back features when predicting either individual items (questions) within the PHQ-8 scale, or predicting overall depression severity as given by the overall PHQ-8 score. We modelled each feature separately, and included age, gender and personal device setup (encoded as dummy variables) as covariates. We ranked feature importance in predicting each item, or the overall score using the R$^{2}$ value of each model. 

\begin{table}[t!]
\caption{Top features for predicting either a single PHQ-8 item or total PHQ-8 score}
\label{tab:feature-ranks}
\begin{tabular}{@{}lrrr@{}}
\toprule
\textbf{Item} & \textbf{Feature} & \textbf{R$^{2}$} & \textbf{Beta}\\ \midrule
Item \#1 & Loudness Peaks Per Sec. (eGeMAPS) & .051 & -.054\\
Item \#2 & Loudness Peaks Per Sec. (eGeMAPS) & .039 & -.049\\ \midrule
Item \#3 & 1-Back Reaction Time & .034 & .089\\
Item \#4 & 1-Back False Positive & .049 & .096\\ \midrule
Item \#5 & CV Spectral Flux Voiced (eGeMAPS) & .052 & .001\\
Item \#6 & Syllable Rate (Praat) & .042 & .001\\ \midrule
Item \#7 & Loudness Peaks Per Sec. & .034 & -.073\\
Item \#8 & overall n-Back Precision & .033 & .055\\ \midrule
Total & Loudness Peaks Per Sec. & .049 & -.322\\ \bottomrule
\end{tabular}
\end{table}

\section{Results and Discussion}
We use the area under the ROC curve (ROC-AUC) as the performance metric to compare the models. We see that the performance of the unimodal models on the training set is well-above chance level on average (Table~\ref{tab:cv_performance}). This performance demonstrates that both speech and n-Back features contain predictive information about the corresponding PHQ-8 score. In addition, the average performance of the Multimodal model is higher than all unimodal models (Table~\ref{tab:cv_performance}), suggesting that the two modalities contain complementary PHQ-8 information.  

To validate these results, we tested the Speech, n-Back and Multimodal models on the test set, which was not touched during model training and hyperparameter calibration. The performance of all models on the test set is qualitatively similar to the training set, with the Multimodal model outperforming the unimodal ones (Figure~\ref{fig:test_roc}). 



Our linear regression analysis offers insights into why the fusion of our Speech and n-Back features improves results. When comparing the top-ranking features for each PHQ-8 item (Table~\ref{tab:feature-ranks}), we can see that the items relating to anhedonia (Item \#1), depressed mood (Item \#2), change in appetite (Item \#5), feelings of worthlessness (Item \#6) and problems concentrating (Item \#7) all returned a speech feature as the top-ranked feature. While the top feature for the somatic (Item \#3 and \#4) and psychomotor (Item \#8) items were from the n-Back. Interestingly, n-Back features are ranked in the top three positions for the somatic and psychomotor items highlighting the strength of this task in capturing these changes. 

\section{Conclusion}
Digital phenotyping offers the chance to aid depression diagnosis and management by providing objective information based on cognitive, physiological and behavioural cues. The results presented in this paper demonstrate that speech features, and metrics derived from n-Back Task performance offer complementary information when predicting depression. This is shown in two ways: by the increase in predictive performance when fusing both modalities, and feature space analysis, indicating that features from the different modalities strongly align to different items within the PHQ-8 domain. To the best of the authors' knowledge, this is the first time such a result has been shown. Future work will focus on repeating the fusion results in more complex models and exploring how the addition of facial information changes the feature space dynamics.

\newpage
\bibliographystyle{IEEEtran}



\end{document}